\documentclass[prl,twocolumn,showpacs,amsmath,amssymb,floatfix]{revtex4}
\usepackage{graphicx}
\usepackage{epsfig}
\usepackage{dcolumn}
\usepackage{bm}

\def\(({\left(}
\def\)){\right)}                       
\def\[[{\left[}
\def\]]{\right]}    

\newcommand{\be}{\begin{equation}}
\newcommand{\ee}{\end{equation}}
\newcommand{\bea}{\begin{eqnarray}}
\newcommand{\eea}{\end{eqnarray}}

\newcommand{\<}{\langle}
\renewcommand{\>}{\rangle}

\begin{document}
   
\title{Glassy properties of the Kawasaki dynamics of two-dimensional
  ferromagnets} \author{Florent Krz\c{a}ka{\l}a} \affiliation{Dipartimento di
  Fisica, INFM and SMC, Universit\`a di
  Roma ``La Sapienza'', P.~A.~Moro 2, 00185 Roma, Italy\\
  Laboratoire P.C.T., UMR CNRS 7083, ESPCI, 10 rue vauquelin, 75005 Paris,
  France} \date{\today}

\begin{abstract}
  We study numerically the Kawasaki dynamics of the $2d$ Ising model.  At
  large time we recover the coarsening growth as $l_c(t) \propto t^{1/3}$. At
  shorter time however, the system enters a metastable glassy regime that
  displays an extremely slow growth and non-trivial violations of the
  fluctuation dissipation theorem similar to those observed in spin glasses:
  this is one of the simplest system where such violations occur. We also
  consider Potts models, where a similar behavior is observed, and the model
  of Shore and Sethna where the domain growth is also slow, but where
  violations of the fluctuation dissipation theorem are trivial. We finally
  comment on these violations in the context of activated coarsening, and on
  similarities and differences with the glass transition phenomenology.
\end{abstract}
\maketitle

\vspace{-0.5cm}
Last decades have witnessed many progresses in the study of out-of-equilibrium
dynamics and coarsening, or domain growth, processes~\cite{Bray94}. In many
simple models, the dynamics after a quench to the ordered phase is understood
in terms of domains growing with time as $t^{1/2}$ with a Glauber (or non
conserved magnetization) dynamics, and as $t^{1/3}$ with
Kawasaki~\cite{Kawasaki66} (or conserved magnetization)
dynamics~\cite{Bray94}. The $2d$ Ising model plays an important role in the
field as one of the simplest models for coarsening and phase separation.
Recently, motivated by spin glass theory, many studies tried to go beyond
these considerations to investigate two-time quantities such as correlation
and response functions, the relation between them, and the way the dynamic
Fluctuation-Dissipation Theorem (FDT) is violated in the out-of-equilibrium
regime~\cite{LeticiaJorge,FDTreview,Silvio}.  A generalization of the FDT can
be formulated and amounts, within the mean field theory of aging, to the
introduction of an effective temperature $T_{eff}$; it is widely accepted that
$T_{eff}=\infty$ for coarsening process~\cite{Peliti}, as confirmed by
simulations in the Ising model with Glauber dynamics~\cite{Barrat98}.  In
glasses and spin glasses however, violations of the FDT are highly non
trivial~\cite{LeticiaJorge,FDTreview,Fede}; finding lattice models with such
properties has recently triggered a lot of attention~\cite{Andrea,Buhot}.

In this letter, we study the $2d$ Kawasaki Ising
model~\cite{Kawasaki66}.  Even if the domains grow as $t^{1/3}$ at
large time~\cite{Huse86,AmarSullivan88}, we feel
interesting to study the short time behavior as pre-asymptotic regimes
are already non trivial in the $1d$
chain~\cite{GodrecheLuck2003,GodrecheKrzakala04}. We will demonstrate
the existence of a glassy early regime, where domain growth is
extremely slow, that displays a violation of the FDT similar to those
of more complex frustrated models such as spin glasses or glass
models. At large time we recover the Glauber-like behavior.  The same
phenomena arises with Potts variables but not in the model of Shore
and Sethna~\cite{Sethna}, although it displays a logarithmic
growth. We finally discuss our results in the context of glassy dynamics.

\paragraph*{Model and methods ---} 
We consider the Ising model on a $2d~L^2$ lattice with Hamiltonian $H = -
\substack{ \sum\\<i,j>} S_i S_j$, whose critical temperature is $T_c \approx
2.27$. To increase the efficiency of the Monte-Carlo (MC) simulations, we use
a few tricks: first, we use the method described in~\cite{GodrecheKrzakala04},
a generalization to Kawasaki dynamics of an algorithm~\cite{Chatelain,Fede}
allowing a computation of the linear response to a field without physically
adding a magnetic field in the simulation. Many two-time quantities for
different waiting times were thus computed in the same run.  Secondly, we
speed up simulations using the continuous time method of~\cite{BOOK}: instead
of choosing at random a pair of neighboring spins and to try to exchange them
at each MC step, we keep track of all the $N_b$ broken links in memory, choose
one at random, increase the clock by the time needed for the system to find it
(a Poissonian variable with mean $2L^2/N_b$; time units being in MC steps per
site) and finally update the value of the corresponding pair of spins using
the Heat Bath method.  Since both tricks can be used at the same time, our
approach allows a very good numerical determination of correlation and
response functions at long time, even for large ($L \approx 10^3$) systems.
Finally, we determined the coarsening length $l_c(t)$ by computing the excess
energy~\cite{AmarSullivan88,FisherHuse88,GodrecheKrzakala04} with respect to
equilibrium $l_c(t) = - E_{eq} / \((E(t) - E_{eq} \))$ (where $E_{eq}$ is
given by the exact solution~\cite{Wu}; with this definition $l_c(t)$ is $O(1)$
in the high temperature phase) and checked that the length defined as the
first zero of the spacial correlation function~\cite{Huse86} gives similar
results.

\paragraph*{Growing length beyond the $t^{1/3}$ law ---}
Let us first consider in details the domain growth process. It is now
clear that it behaves as $t^{1/3}$ at late time but first numerical
studies failed to find this exponent and, after initial claims in
favor of a logarithmic coarsening~\cite{Mat}, Huse stressed the
importance of activated phenomena and energy barriers~\cite{Huse86}:
due to the spin exchange dynamics, some moves need activation before
happening, which creates different time scales. For instance, zero
temperature simulations converge toward highly non trivial
configurations far from equilibrium~\cite{Mat}, where they remain
blocked forever and therefore, since first excitations cost an
activation energy $\delta E=4$, any finite temperature simulations for
a time shorter than $\tau_1=e^{4\beta}$ will behave as those at
$T=0$. Going at time larger than $\tau_1$ is not even enough to access
the late time regime and in fact it is only when $t \gg
\tau_2=e^{8\beta}$\cite{Huse86,AmarSullivan88} that the
system reaches the proper power-law behavior.

\begin{figure}
\centerline{\hbox{\epsfig{figure=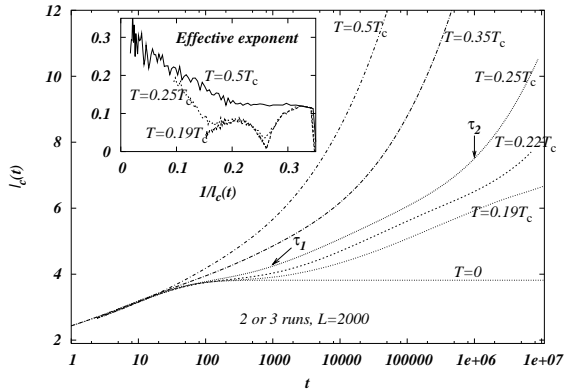,width=7.6cm}}}
\caption{Coarsening length versus time at low $T$ in a log-linear
  plot. After a finite time ($O(10^2)$) the dynamics reaches the zero
  temperature plateau. After an time $\tau_1 \propto e^{4\beta}$, it enters a
  regime with a slow domain growth and beyond time $\tau_2 \propto e^{8\beta}$
  eventually enters the $t^{1/3}$ Kawasaki scaling ($\tau_1$ and $\tau_2$ are
  indicated for $T=0.25$ on the picture). In inset: effective growth exponent
  versus $l_c(t)^{-1}$, that converges to $1/3$.}
\label{Grows}
\end{figure}
Doing long quenches at many low temperatures, we investigate precisely
the behavior of $l_c(t)$ and summarized our results in FIG.\ref{Grows}
where three different regimes for the dynamics after a quench to $T
\ll T_c$ can be observed: (A) {\it the Zero Temperature regime:} after
a finite time, one reaches the plateau characteristic of the $T=0$
dynamics, (B) {\it the Glassy regime:} after an activation time
$\tau_1 \propto e^{4 \beta}$, the dynamics leaves the plateau and
enters a regime with a very slow domain growth. Actually, $l_c(t)$ seems
even sub-logarithmic at large time, suggesting, as first intuited
by~\cite{Huse86}, that $l_c(t)$ saturates and reaches a second plateau
at very large $t$ (this is however quite invisible at the time scales
considered here), and (C) {\it the Asymptotic regime:} after a time
$\tau_2 \propto e^{8 \beta}$, the dynamics enters the canonical regime
where $l_c(t) \approx A + B t^{1/3}$, as can be seen from the behavior
of its effective growth exponent (inset in FIG.\ref{Grows}).

This glassy regime, whose name will be justified later, thus corresponds to a
metastable state with lifetime $O(\tau_2)$ that lasts extremely long at low
temperature. That $l_c(t)$ seems to saturate indicates that the system is
growing small organized domains that then remain blocked because further
growth would need a move costing $\delta E = 8$ (i.e. flipping a spin on
domain edges). This looks very similar to the polycrystalline picture (a
collection of small crystalline zone) observed in the early stages of the
dynamics of some model glass~\cite{Andrea}. We also noticed that the way the
system escapes from these metastable states may involve cooperative phenomena
and non just a single spin flip of energy $8$: modifying the dynamics in such
a way that only the moves that cost $\delta E \le 4$ are allowed, we found
that the dynamics still converge to the $t^{1/3}$ regime after a time,
although a bit larger, still scaling as $e^{8\beta}$. To emphasize the
differences between the dynamics in this glassy regime and the late time
coarsening, we turn now to FDT violations.

\paragraph*{Violation of the FDT ---}  
\begin{figure}
\centerline{\hbox{\epsfig{figure=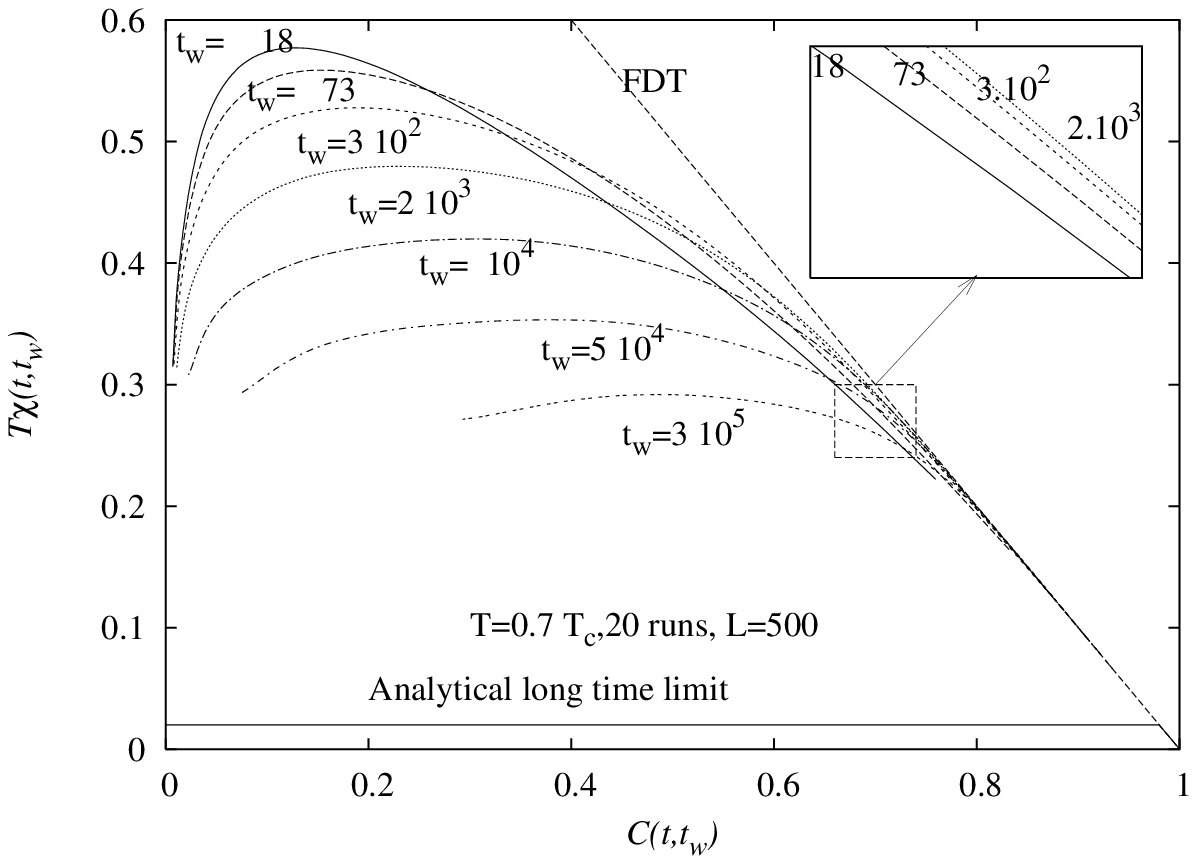,width=7.6cm}}}
\centerline{\hbox{\epsfig{figure=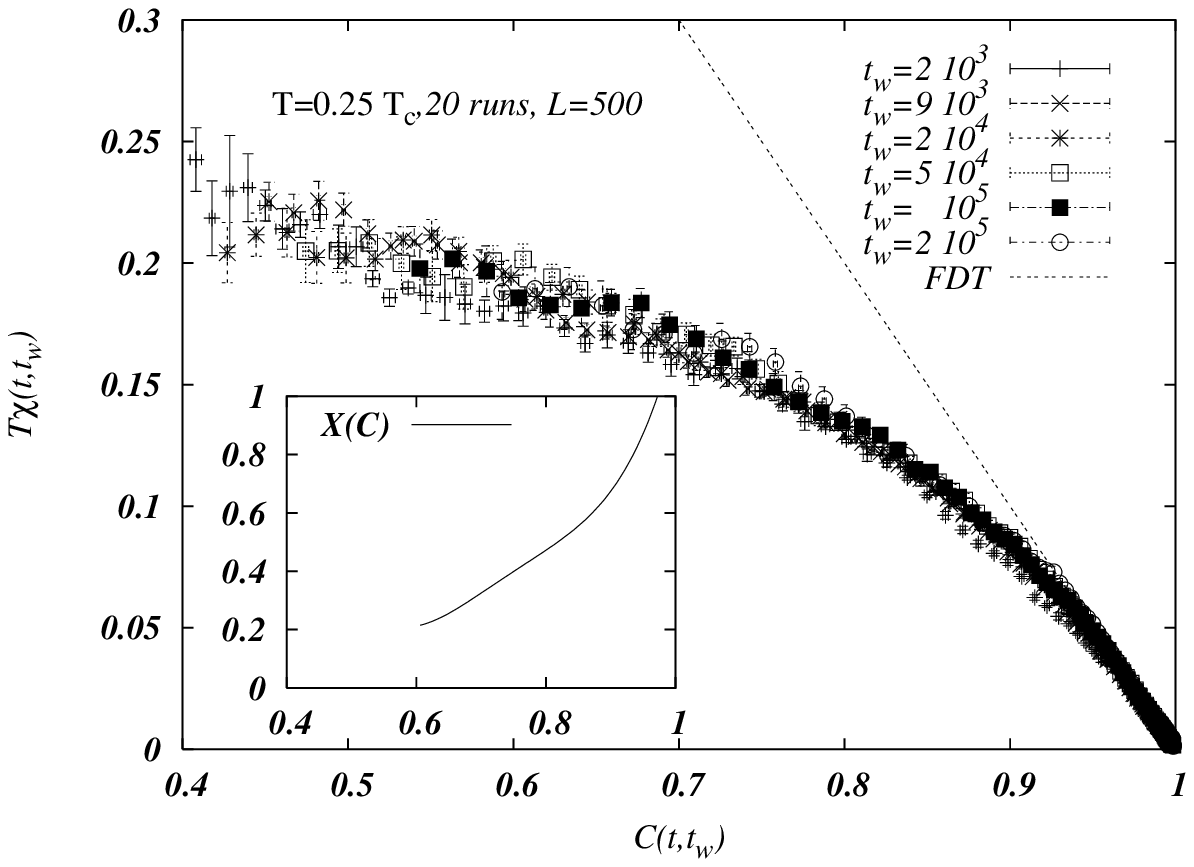,width=7.6cm}}}
\caption{Up (2.a) : FDR plot at $T=0.7 T_c$. At long time, curves converge
  slowly to the expected limit, as for Glauber dynamics~\cite{Barrat98}.
  However, at short time, curves are shifting up as $t_w$ increases, like in
  spin glass models. Down (2.b): FDT for $T=0.25 T_c$, in the glassy regime
  only; curves are superposing very well on two decades in $t_w$, and
  correspond to a non trivial quasi-equilibrium regime. In inset the
  corresponding $X(C)$.}
\label{FDT_g}
\end{figure}
Studies of the {\em Fluctuation-Dissipation Ratio}
(FDR)~\cite{LeticiaJorge,FDTreview} are based on the comparison of how
spontaneous and induced fluctuations do relax: one measures the
two-time correlation function $C(t,s)=\<S(t)S(s)\>$ and the associated
response to a field function $R(t,s)=\partial \< S(t) \> /\partial
h(s)$, and defines the FDR $X(t,s)$ through the formula
%
$T\,R(t,s) = X(t,s)\;\partial_s C(t,s).$
%
At equilibrium the FDT holds, thus $X=1$. In the off-equilibrium regime
however, the FDT is violated. In the large time limit ($s,t\!\to\!\infty$ with
$C(t,s)\!\to\!q$) the FDR $X(t,s)$ converges to a limiting function $X(q)$
whose physical meaning has been derived within the mean field theory of aging,
where it has been shown~\cite{Silvio} that it can be computed using the
overlap pdf $P(q)$ in the threshold states (the states reached by the dynamics
on very large time~\cite{FDTreview}) using $X(q) = x(q) \equiv \int_0^q P(q')
dq'$. The interpretation of $X(q)$ in finite $d$ models, the meaning or even
the existence of a unique effective temperature $T_{eff} = T/X$ play a central
role in present research in off-equilibrium physics~\cite{FDTreview}.
Experiments and simulations consider integrated quantities, in which
case one gets
\be 
\chi(t,t_w) = \frac{1-C(t,t_w)}{T_{eff}},
\label{eq_r2}
\ee where $\chi(t,t_w)$ is the linear susceptibility at time $t$ that
results from a magnetic field switched on from times $t_w$ to $t$. A
popular presentation of such data~\cite{LeticiaJorge} is a plot of
$T\chi(t,t_w)$ versus the $C(t,t_w)$: one finds that FDT holds at
short time, when $C(t,t_w)$ is large (in the so-called
quasi-equilibrium regime) so that $T_{eff}=T$ until $C(t,t_w)$ drops
and reaches a given value $q_{EA}$~\cite{FDTreview} (equal to the
exactly known equilibrium magnetization in the Ising model with
Glauber dynamics) and FDT is violated for $C(t,t_w)<q_{EA}$ where
the introduction of $T_{eff}>T$ is thus needed.

It is widely believed that any coarsening processes is characterized by
$T_{eff}=\infty$. This is quite expected since the $P(q)$ corresponding to
ferromagnetic equilibrium is trivial.  Here, however, there is the early
regime where the dynamics is reaching metastable states far from the
equilibrium ones, so, inspired by the mean field interpretation of FDT
violations, we expect that in the glassy regime they are not ruled by the
equilibrium ferromagnetic $P(q)$ but by the glassy metastable states so that
the FDR is non trivial.  We present our data for $T=0.7 T_c$ in
FIG.\ref{FDT_g}.a: at large time, the curves converge very slowly toward the
expected limit, approaching it downward as in the Glauber case (a direct
comparison with~\cite{Barrat98,Fede} is instructive). This excess response
signal can be understood by noticing that spins on domain walls have low local
fields and thus participate a lot to the global magnetic response.  At short
time $l_c(t)$ is quite low so there is a large excess density of domains that
eventually decreases at larger time as domains annihilate: this gives rise to
the bump in the FDT plot and explain the slow downward convergence of the
susceptibility as $t_w$ increases. All this is similar with what happens with
Glauber dynamics~\cite{Barrat98}.

However, a new phenomenon arises at shorter time: in FIG.\ref{FDT_g}.a curves
are first drifting up as the waiting time $t_w$ increases. This is in {\it
  sharp} contrast with what is observed with Glauber dynamics and quite
similar to spin glass behavior ~\cite{Fede}: in this regime the system is
somehow creating more response as $t_w$ increases. These phenomena arise for
time scale $O(\tau_2) \approx 10^2$; this suggests that, staying long enough
in the glassy regime, by working at a lower temperature, we should observe
clearer FDT violations. This is indeed the case (see FIG.\ref{FDT_g}.b) at
$T=0.25 T_c$ (where $\tau_2 \approx 10^6$) where we observe a superposition of
FDR curves on more than two decades and obtain a plot very similar to those of
complex systems as glasses and spin glasses.  This shows that, between times
$\tau_1$ and $\tau_2$, the dynamics of the Kawasaki model in the glassy regime
is very sensitive to the metastable states and, as expected, exhibits a non
trivial FDR. It is interesting that ideas arising in mean field disordered
complex systems are realized in a certain time range in the $2d$ Ising model,
which as opposed to other simple models where FDT violations where
observed~\cite{Andrea,Buhot,BB}, is not critical or close to a critical point,
not disordered and not even statically frustrated.  Despite that, it has the
striking properties of behaving like a ferromagnet at long time but as a spin
glass at short time, suggesting that any coarsening model with similar
activated regimes could behave like a glass at short times. 

\paragraph*{Logarithmic coarsening ---} 
\begin{figure}
\centerline{\hbox{\epsfig{figure=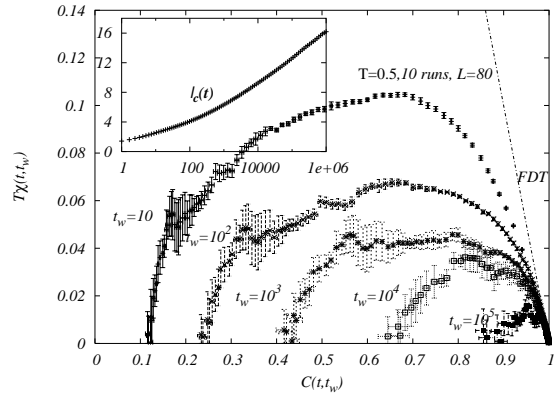,width=7.6cm}}}
\caption{FDR plot for the $3d$ Shore-Sethna model with competitive
  ferromagnetic $(J=1)$ $1^{st}$ neighbor and anti-ferromagnetic $(J=-1/6)$
  $2^{nd}$ neighbor interactions. We observed a very slow domain growth
  (inset) but a trivial FDR.}
\label{FDT_Sethna}
\end{figure}
To investigate the generality of our results, we followed Shore and Sethna
(SS)\cite{Sethna} who introduced a $3d$ Ising model with ferromagnetic first
neighbors but weak anti-ferromagnetic second neighbors interactions, and
argued that it has barriers growing with the size $l_c(t)$ of the
ferromagnetic domains, thus leading to an activated logarithmic
coarsening~\cite{Mat}. We repeated our FDT study in the range of temperature
and parameters where SS claimed to observe a logarithmic growth and computed
$l_c(t)$ using the same definition as~\cite{Sethna}. We present our results in
FIG.\ref{FDT_Sethna}: although we find data compatible with a logarithmic
$l_c(t)$ at large time, $\chi$ is very small, the FDR seems trivial and all
our FDT plots look like what is obtained for usual coarsening
process~\cite{Barrat98}; in particular all curves are shifting downward as
$t_w$ increases, as opposed to FIG.\ref{FDT_g}.a. This demonstrates that it is
{\it not} just the slowly growing length that creates the non trivial FDR in
FIG.\ref{FDT_g}. There, the dynamics clearly differs from usual coarsening,
even logarithmic, and reaches some non trivial quasi-equilibrium states that
seem stable until times of $O(\tau_2)$; this is in contrast with the
logarithmic coarsening of the SS model which is just going {\it slowly} toward
the ferromagnetic state, and where no such equilibration in a metastable state
could be defined.

\begin{figure}
\centerline{\hbox{\epsfig{figure=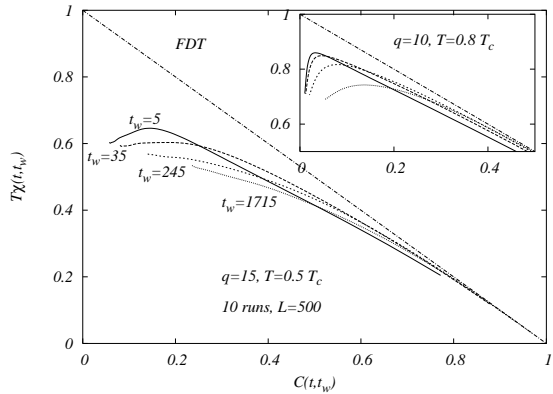,width=7.6cm}}}
\caption{FD in the $q$-Potts model for $q=10$ (inset, $T=0.8
  T_c$) and $q=15$ at $T=0.5 T_c$. As for Ising, one sees a crossover from the
  early regime to coarsening for $t >> \tau_2 \approx O(10)$.}
\label{FDT_Potts}
\end{figure}

\paragraph*{Potts variables and glasses ---} 
The present study is easily extended to the q-states Potts model~\cite{Wu}
(where $T_c = q/\log{(1+\sqrt{q})}$) and we checked that a similar
phenomenology for the FDR and the slow dynamics arises: the same
activated processes give rise to glassy behavior (see FIG.\ref{FDT_Potts}).
This is interesting as Potts models have a $1^{st}$ order transition for
$q>4$, hence we obtain a model with many features of the glass phenomenology
such as FDT violations, melting transition and metastable states. Moreover,
like other models with much more complicated Hamiltonians recently developed
to study analytically the glass transition~\cite{GlassLattice}, our Ising or
Potts magnets with Kawasaki dynamics have a non trivial mean-field limit on
Bethe lattices, where they display a spin glass behavior (within the cavity
method~\cite{SG}, the spin glass solution applies to ferromagnetic couplings
when the magnetization is fixed to zero)
so we may ask if this model is pertinent as a model glass and how it compares
with~\cite{GlassLattice} ? As in any realistic models, the low $T$ phase
eventually crystalizes at large time once the free energy barriers to crystal
nucleation are crossed (if the nucleation time is shorter than the
equilibration time of the metastable glass state, the later cannot be
thermodynamically defined: this is the solution Kautzmann originally proposed
to its own paradox~\cite{Kautz}).  A lifetime of $O(e^{8 \beta})$ being not
very large close to $T_c$, it is quite easy in this model to overcome these
barriers in a cooling experiment so that no supercooled state and no glass
transition can be observed (this is also what happens in the model
of~\cite{Andrea} obtained with a very frustrated plaquette Hamiltonian): here,
it is only by {\it quenching} the system fast enough to low $T$ that it shows
its glassy nature. Precautions are thus needed when using mean field methods;
it suggests however that a similar model with barriers large enough not be
overcome in the time scale of the simulation would be a good model; this is
probably what happens in models~\cite{GlassLattice} which are believed
to be good glass formers.

\paragraph*{Discussion ---} 
We studied the off-equilibrium dynamics of $2d$ Kawasaki magnets and
demonstrated the existence of a early extremly slow glassy regime with a non
trivial violation of the FDT similar to those of spin glasses. We observed the
same phenomenology with Potts variables but not in the SS model, pointing out
the difference between this dynamical behavior and the usual coarsening,
however slow it may be. We believe that this behavior is quite generic and
should be observed (as it has already been in some specific
cases~\cite{Andrea,Buhot}) in the early stage of other coarsening model
(provided that they share similar activation properties) or in some lattice
model for glasses. Since non trivial FDT violations have been observed in many
systems~\cite{FDTreview}, it is interesting that a simple model such as the
$2d$ Ising ferromagnet, whose statics solution is analytically known for
decades, has a dynamics that already shares many properties with those of much
more complex materials: very few ingredients are needed for that. Finally our
results show, as was pointed out in~\cite{BB}, that great cares should
be taken when using dynamical simulations to probe the statics.

\paragraph*{Acknowledgment ---}
I warmly thank A.~Cavagna, F.~Ricci-Tersinghi and T.~J\"org for
discussions and acknowledge support from European Community's Human
Potential program HPRN-CT-2002-00319 (STIPCO)

\end{document}